# On possible spin injection at non-ideal Schottky contacts

Dean Korošak[a*] and Bruno Cvikl[a,b]

[a]Chair for Applied Physics, Faculty of Civil Engineering,

University of Maribor, Smetanova 17, 2000 Maribor

and

[b] "J. Stefan" Institute, Jamova 19, Ljubljana

Abstract

The role of the tunneling mechanisms in metal-disordered layer-semiconductor structure under spin injection at the interface is investigated. The non-ideal metal-semiconductor structure as prepared by ionized cluster beam deposition is considered, and it is shown that the depletion region of the semiconductor can be tailored to include a suitably heavily doped region near the interface. The tunneling is described within a simplified model in which the expression for the interface resistance of the metal-disordered layer-semiconductor structure is obtained. It is argued that in the case of ionized cluster beam deposited non-ideal Schottky structure a significant spin injection is achieved.

---

[*] Corresponding author: fax +386 2 2524 179, email: dean.korosak@uni-mb.si

The physics of metal-semiconductor contacts has a long and rich history [1,2] throughout which a number of theoretical models were proposed and controversial issues raised in order to uniquely describe the various metal-semiconductor systems and their physical properties. Metal-semiconductor interfaces are crucially important to modern micro- and nanoelectronic devices, and together with the spatial diminishing of their active areas are the details of interface electronic structure becoming of the paramount importance for the control of the charge carrier transport through the interface. Modern experimental techniques [3] already allow us to investigate the properties of the interface region with atomic resolution, but the unifying theoretical model does still not exist which is especially true for the *non-ideal* interfaces where the presence of foreign atoms or intentionally inserted layer or various defects are causing deviations from the ideal cases [4,5].

Metal-semiconductor interfaces again attracted considerable attention in the last few years [6,7,8,9] when it was suggested that the high values of spin polarization can be obtained in spin dependent tunneling, and that the Schottky barrier would resolve the conductance mismatch problem [10] serving as a resistive interlayer [11,12]. Recently [13], the predominately role of tunneling transport mechanism in spin injection at Fe/AlGaAs Schottky contact has been experimentally verified.

Closely connected to the questions of the microscopic properties of the m/s interface is the problem of the excess capacitance [14,15,16] in Schottky junctions. We have been able to show [5] that the observed excess capacitances in non-ideal Schottky junctions arise on account of the bias dependent localized charge density induced at the appropriate interface in the structure.

In what follows the possibility of using a non-ideal Schottky contact with a disordered interlayer between the metal and the semiconductor as a spin injector is investigated. Specifically, we explore these possibilities in the special cases of devices prepared by ionized cluster beam deposition [5]. The structure of ICB deposited metal-semiconductor contact in our view consists of the homogenous metal overlayer, the regular bulk semiconductor and the interlayer coupling the metal and the semiconductor. The width of the interlayer is given by the penetrating length $L$ of the accelerated metal atoms and clusters while its structure determine incorporated metal atoms in the host semiconductor lattice. Albrecht and Smith [6] have shown that the detrimental influence of the barrier depletion region can be reduced if the barrier height is effectively lowered due to the heavily doped region near the interface. They estimated that the interface resistance of the spin majority carriers should be of the order of $10^{-3}$ $\Omega cm^2$, and the barrier should be lowered to about 0,2 eV. The analysis of the capacitance-voltage characteristics in non-ideal Schottky structures [5] yielded the following expression for the potential in the depletion region of the device which is divided into the thin heavily doped disordered region ($L>x>0$) and the regular semiconductor ($w>x>L$):

$$V(x) = \begin{cases} \phi_{b,0} - en_1(x^2 - 2Lx)/2\varepsilon_d - en_2 x(w-x)/\varepsilon_d - Q_{dsc}(U) x/\varepsilon_d, L > x > 0 \\ en_2(w-x)^2/2\varepsilon_{sc} + U, w \geq x > L \end{cases} \quad (1)$$

Here, $\phi_{b,0}$ is the barrier height at the metal, $n_{1,2}$ are the doping densities and $\varepsilon_{d,sc}$ the dielectric permitivities in the two regions, and $U$ is the applied bias. The width of the depletion layer is [5]:

$$w = \sqrt{L^2(\varepsilon_d/\varepsilon_{sc} + n_1/n_2)(\varepsilon_d/\varepsilon_{sc}) + 2\varepsilon_{sc}(\phi_{b,0} - U - Q(U)L/\varepsilon_d)/en_2} \quad (2)$$

The specific of the model is the introduction of the bias dependent interface charge density $Q_{dsc}$ positioned at the penetration length of the metal atoms $L$. In figure 1 we show the zero bias calculated potential in silicon for the following set of parameters: $\phi_{b,0}$=0,6 eV, $n_1$=$n_2$=$10^{16}$ cm$^{-3}$, $L$=2,5 nm. These parameters are typical for the metal/silicon non-ideal interface as deduced from analysis of the capacitance-voltage characteristics of Ag/Si, Pb/Si, Al/Si, TiSi$_2$/Si, Mo/Si Schottky structures [5]. The three curves presented correspond to $Q_{dsc}$=0,01 C/m$^2$, 0,02 C/m$^2$ and 0,025 C/m$^2$ from top to bottom. The values chosen here are an order of magnitude larger than those deduced from experimental data which were typically in the range from 0,002 to 0,007 C/m$^2$ [5], and illustrate the effect of the interface charge density produced in the heavily damaged semiconductor layer. Even without the significant additional doping in the disordered layer, large enough interface charge density induced at the disordered layer/semiconductor interface the depletion layer yield the desired modification of the depletion layer, sharply decreasing the thickness of the Schottky barrier.

We next consider the model for the interface resistance of the non-ideal Schottky contact. The spin injection into semiconductors is most efficient if the polarized electrode (i.e. ferromagnet metal) and the bulk semiconductor are decoupled with the thin barrier through which electrons can tunnel. In our case is the decoupling layer the disordered region characterized by its width $L$ and the density of states $\rho_{dsc}$. The tunneling current density is [17]:

$$J = \frac{4\pi e}{\hbar} \sum_{k_t} \int \left|M_{m,dsc}(E)\right|^2 \rho_m(E) \rho_{dsc}(E) \left(f_m(E - E_{Fd}) - f_{dsc}(E)\right) dE , \qquad (2)$$

where $M_{m,dsc}$ is the tunneling matrix element, $\rho_m$ density of states in the metal, $E_{Fd}$ is the difference in Fermi energy across the disordered layer, and $f_{m,dsc}$ are the Fermi-Dirac functions. The sum goes over the transverse wave-vector which is conserved in the tunneling. For the calculation of the tunneling matrix element we follow the simplified Bratkovski's model presented in [18] by neglecting the dependence of the tunneling matrix element on the transverse wave-vector altogether. The obtained result for the tunneling current density is:

$$J_t = \frac{4e\hbar m D_{dsc} E_{Fd} \gamma_d}{\left((m\gamma_{sc})^2 + (m_t k_F)^2\right)} k_F^3 \exp(-L\gamma_s), \tag{3}$$

where the $D_{dsc}$ is the interface density of states, $\gamma_{sc} = \frac{1}{\hbar}\sqrt{2m_{sc}(E_C - E_B)}$ and $\gamma_d = \frac{1}{\hbar}\sqrt{2m_t(E_{C,d} - E_{B,d})}$. Here, $E_C$ ($E_{C,d}$) is the bottom of the conduction band, $E_B$ ($E_{B,d}$) is the Fermi level pinning position [4] in the bulk semiconductor (disordered layer). $m$, $m_{sc}$, $m_t$ are the effective electron mass in the metal, semiconductor and disordered layer whereas $\gamma_s$ is the function of both these values: $\gamma_s = (\gamma_{sc} - \gamma_s)/\ln(\gamma_{sc}/\gamma_s)$ [18]. The Fermi wave-vector depends on the density of electrons in the metal as $k_F = \sqrt[3]{3\pi^2 n}$. From the definition of the tunneling resistance and from the admittance analysis of the metal-semiconductor interface [18], the expression for the spin dependent interface resistance yields:

$$R_{t\uparrow} = \frac{\left((m\gamma_{sc})^2 + (m_t k_{F\uparrow})^2\right)}{e\hbar m D_{dsc} \gamma_d k_{F\uparrow}^3 \left(1 + \frac{\varepsilon_d}{e^2 L D_{dsc}}\right)^2} \exp(L\gamma_s), \tag{4}$$

where $\varepsilon_d$, the dielectric permitivity of the disordered layer, is taken to be 2 $\varepsilon_{sc}$. Here, a definition of the interface dielectric constant $\varepsilon_d^{-1} = (\varepsilon_m^{-1} + \varepsilon_{sc}^{-1})/2$ [19] is used in the limit $\varepsilon_m = \infty$ [20, 21] for the metal.

The spin transport through the polarized electrode-semiconductor contact is usually described within the drift-diffusion theory. In order to calculate the spin injection ratio at metal-disordered layer-semiconductor structure we use the model devised by Smith and Silver [7]. There the spin-flip scattering is neglected, and the spin-up and spin-down electrons are treated as separate charge carriers. The spin polarized current density is $J_{\uparrow\downarrow} = \sigma_{\uparrow\downarrow} \partial \mu_{\uparrow\downarrow} / e \partial x$, where $\sigma$ is the metal conductivity of either spin polarization, and $\mu$ is the electrochemical potential. If there is finite interface resistance the electrochemical potential is discontinuous at the interface $J_{\uparrow\downarrow}|_{x=0} = (\mu_{\uparrow\downarrow}^{0+} - \mu_{\uparrow\downarrow}^{0-})/eR_{\uparrow\downarrow}$. The injected current spin polarization is defined as $P = (J_\uparrow - J_\downarrow)/J = 2\beta - 1$. The calculation procedure of the model which we follow here is outlined in detail in ref. [7]. To achieve effective spin injection in the semiconductor, first the non-equilibrium situation must be established at the interface by an electric current. The spin-polarization must then be maintained at the semiconductor side of the contact by decoupling the electrons on both sides of the interface with sufficiently large interface resistance [6,7]. We use equation (4) to estimate the range of attainable interface resistance at spin-polarized electrode/disordered layer/semiconductor structure and then we calculate the current spin polarization. In figure 2 the calculated interface resistance for the two spin channels as a function of density of states at the disorder/regular semiconductor interface is presented. The following set of parameters were used for the calculation: $m=m_e$, $m_{sc}=0,98\ m_e$, $n=6\ 10^{22}$ cm$^{-3}$, $E_c$-$E_B$=$E_{c,d}$-$E_{B,d}$=0,5 eV, $L$=2,5 nm, and $m_t=m_e$. In figure 3 the calculated current density spin polarization is presented. The interface resistances for the spin-up and spin-down channel were calculated using the maximum value for the density of states $D_s = \varepsilon_d / e^2 L$ and the parameters as for the calculation shown in figure 1 except for the tunneling effective electron mass which was $m_t=m_e$ for the upper curve and $m_t=0,7\ m_e$ for the lower curve. The additional parameters for the spin injection ratio calculation were taken the same as those suggested by Silver and Smith [7]: spin polarization in the bulk of the metal was 0,8, the spin diffusion length in metal was 100 nm, and in semiconductor 1 μm. The resitivity of the semiconductor was 1 Ωcm, and the current density through the structure was 1 A/cm$^2$.

From the results shown, we can deduce that the considerable spin injection could be achieved in spin-polarized electrode/disordered layer/semiconductor structure, for interface resistances of the order of $10^{-3}$ Ωcm$^2$ which are achieved using a moderate choice of parameters describing the contact.

Based on the theoretical considerations and results presented here, the non-ideal metal/disordered layer/semiconductor Schottky structure, prepared utilizing the ionized cluster beam deposition, could serve as a potential candidate for the effective Schottky barrier spin injector. The current spin polarization is shown to critically depend on the details of the disordered interlayer the width of which and the properties can be to some extend controlled in the ionized cluster beam deposition experiment. In the present approach these details are incorporated in the model by introducing the tunneling effective electron mass. The selection of the parameters of the model chosen to obtain numerical results as shown in figs. 1-3 was based on analysis of experimentally obtained electrical characteristics of non-ideal Schottky contacts [5].

The efforts for the experimental verification of the theoretical predictions given here are currently underway.

Acknowledgement: The authors would like to thank Professor Alex Bratkovski, Hewlet-Packard Laboratories, Palo Alto USA, for stimulating comments on modeling the tunneling transport processes in spin injection at metal/semiconductor interface.

Figure captions:

Figure 1: The zero bas potential in silicon calculated from equation (1) with the following set of parameters: $\phi_{b,0}$=0,6 eV, $n_1$=$n_2$=$10^{16}$ cm$^{-3}$, $L$=2,5 nm. The three curves presented correspond to $Q_{dsc}$=0,01 C/m$^2$ , 0,02 C/m$^2$ and 0,025 C/m$^2$ from top to bottom respectively.

Figure 2: The calculated interface resistance for the two spin channels as a function of density of states at the disorder/regular semiconductor interface. The following set of parameters were used for the calculation: $m$=$m_e$, $m_{sc}$=0,98 $m_e$, $n$=6 $10^{22}$ cm$^{-3}$, $E_c$-$E_B$=$E_{c,d}$-$E_{B,d}$=0,5 eV, $L$=2,5 nm, and $m_t$=$m_e$.

Figure 3: The calculated current density spin polarization in spin-polarized electrode (FM)/disordered layer/semiconductor (SC) structure is presented. The parameters were from the calculation of interface resistance shown in figure 1 and $D_s = \varepsilon_d / e^2 L$, and the tunneling effective electron mass was $m_t$=$m_e$ for the upper curve and $m_t$=0, 7 $m_e$ for the lower curve. The spin diffusion length in metal was 100 nm, and in semiconductor 1 μm. The resitivity of the semiconductor was 1 Ωcm, and the current density through the structure was 1 A/cm$^2$.

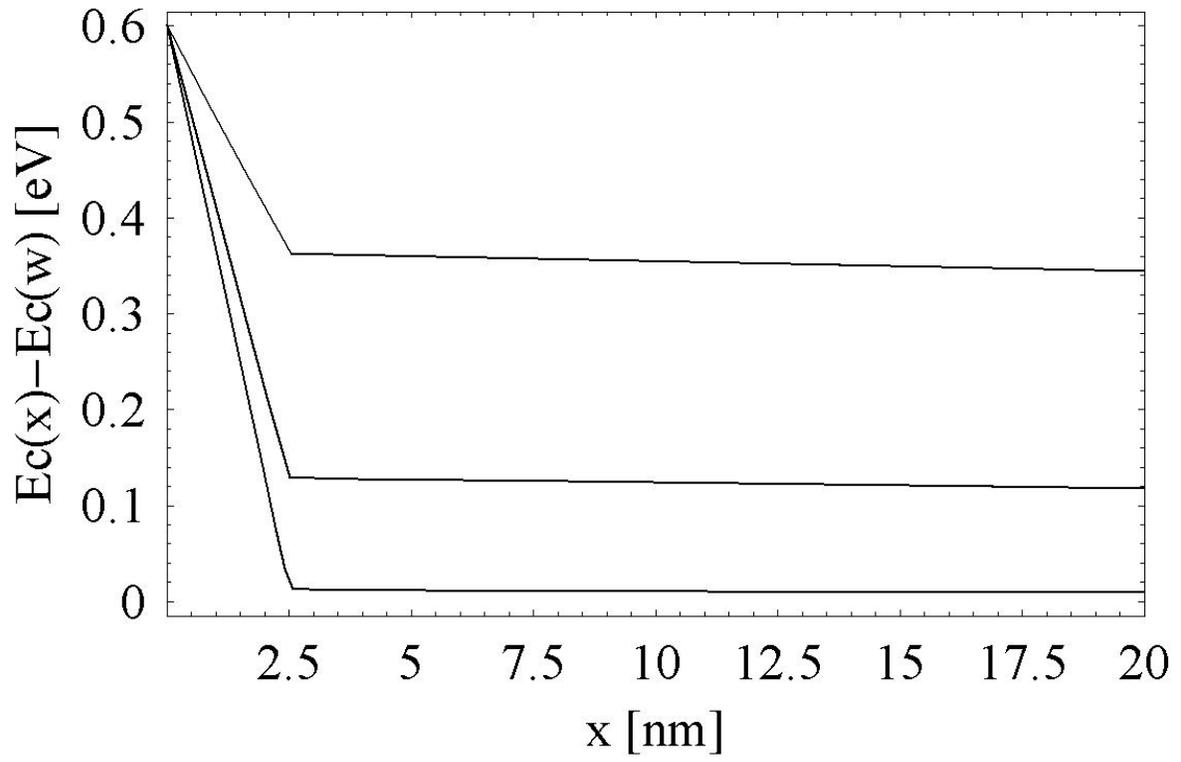

D.Korošak and B. Cvikl, Figure 1

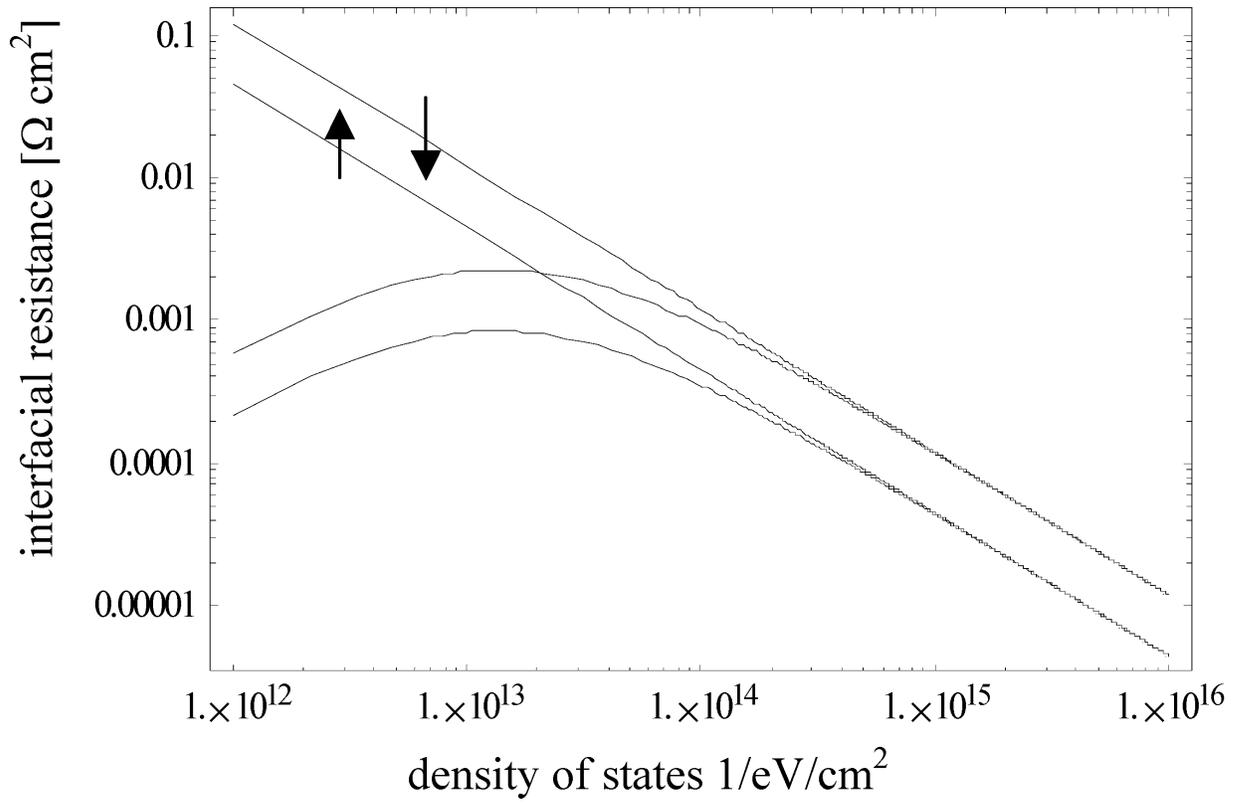

D. Korošak and B. Cvikl, Figure 2

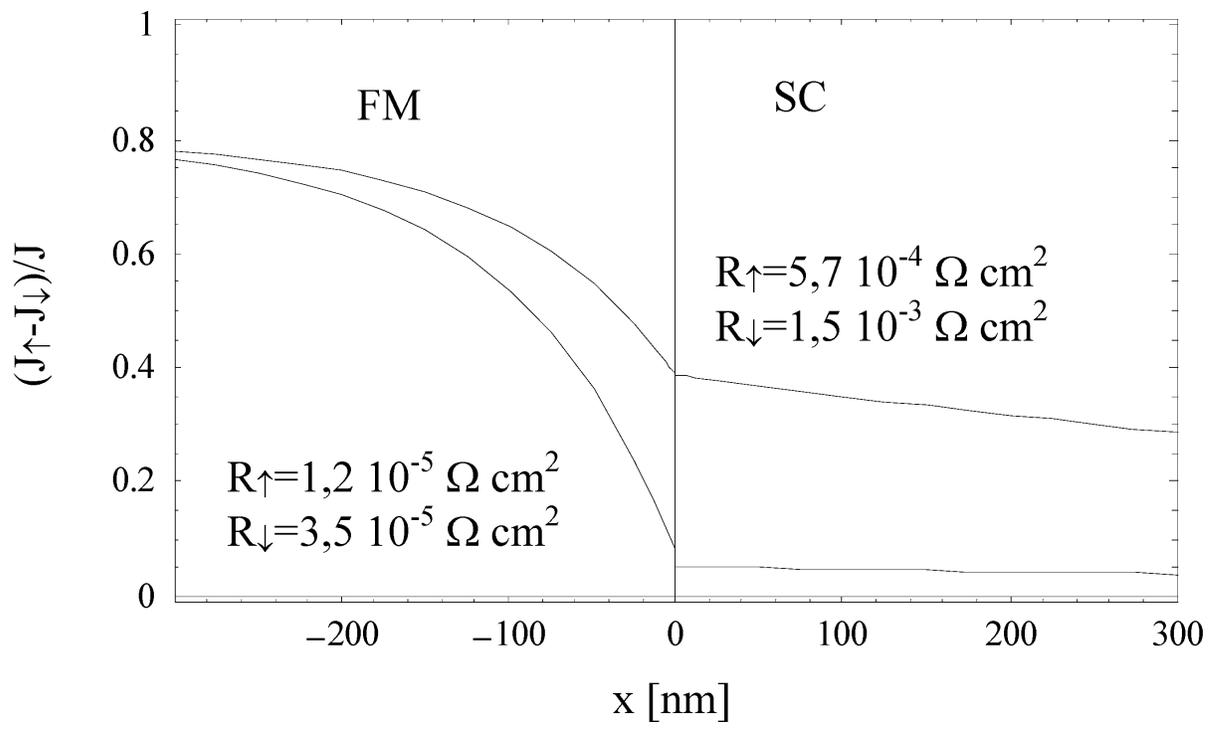

D. Korošak and B. Cvikl, Figure 3